\documentclass[12pt]{article}

\usepackage{scicite}
\usepackage{times}
\usepackage{amsmath}
\usepackage{placeins}
\usepackage{braket}
\usepackage{graphicx}
\usepackage{xcolor}
\usepackage{siunitx}
\usepackage{chapterbib}

\newenvironment{sciabstract}{%
\begin{quote} \bf}
{\end{quote}}

\newcounter{lastnote}

\title{Picosecond Schr\"odinger cat states for ultrafast optical quantum processing}

\author
{Mamoru Endo,$^{1,2\ast}$ Kan Takase,$^{1,2,3}$ Takefumi Nomura,$^{1}$ Tatsuki Sonoyama,$^{1}$\\
Kazuma Takahashi,$^{1}$ Sachiko Takasu,$^{4}$ Daiji Fukuda,$^{4,5}$ Takahiro Kashiwazaki,$^{6}$\\
Asuka Inoue,$^{6}$ Takeshi Umeki,$^{6}$ Peter van Loock,$^{7}$ Petr Marek,$^{8}$ Radim Filip,$^{8}$
\\Warit Asavanant,$^{1,2,3}$ Akira Furusawa$^{1,2,3\ast}$\\
\\
\normalsize{$^{1}$Department of Applied Physics, School of Engineering, The University of Tokyo}\\
\normalsize{7-3-1 Hongo, Bunkyo, Tokyo, 113-8656, Japan}\\
\normalsize{$^{2}$Optical Quantum Computing Research Team, RIKEN Center for Quantum Computing}\\
\normalsize{2-1 Hirosawa, Wako, Saitama 351-0198, Japan}\\
\normalsize{$^{3}$OptQC Inc. }
\normalsize{1-21-7 Nishi-Ikebukuro, Toshima-ku, Tokyo 171-0021, Japan}\\
\normalsize{$^{4}$National Institute of Advanced Industrial Science and Technology}\\
\normalsize{1-1-1 Umezono, Tsukuba, Ibaraki, 305-8563, Japan}\\
\normalsize{$^{5}$AIST-UTokyo Advanced Operando-Measurement Technology Open Innovation Laboratory}\\
\normalsize{1-1-1 Umezono, Tsukuba, Ibaraki, 305-8563, Japan}\\
\normalsize{$^{6}$Device Technology Labs, NTT Inc.}\\
\normalsize{3-1 Morinosato Wakamiya, Atsugi, Kanagawa 243-0198, Japan}\\
\normalsize{$^{7}$Institute of Physics, Johannes Gutenberg-Universit\"{a}t Mainz}\\
\normalsize{Staudingerweg 7, 55128 Mainz, Germany}\\
\normalsize{$^{8}$Department of Optics, Palacky University}\\
\normalsize{17. Listopadu 1192/12, Olomouc 77148, Czech Republic}\\
\\
\normalsize{$^\ast$To whom correspondence should be addressed;}\\
\normalsize{E-mail: endo@ap.t.u-tokyo.ac.jp (M.E.); akiraf@ap.t.u-tokyo.ac.jp (A.F.).}
}

\date{}

\begin{document}

\baselineskip24pt

\maketitle

\begin{sciabstract}
Non-Gaussian states are essential resources for universal, fault-tolerant optical quantum computing, but their generation rate remains limited by low heralding probabilities and operation in nanosecond temporal modes. Here, we demonstrate multi-photon generalized photon subtraction in picosecond optical wave packets, establishing the state-generation capability required for high-rate operation by addressing the temporal-mode bottleneck that has constrained the achievable rate. Two interfering ultrashort squeezed vacua are heralded by photon-number-resolving detection with a high-speed transition-edge sensor and characterized by pulsed homodyne detection matched to 10-ps temporal modes at a 5-MHz pump repetition rate. We reconstruct Wigner functions without loss correction that exhibit up to four distinct negative regions for four-photon heralding, together with an effective cat-state amplitude of $\alpha_{\mathrm{eff}} = 1.69$. This amplitude approaches the range of practical relevance for fault-tolerant cat-code architectures and for adaptive breeding toward logical-qubit generation, while the picosecond temporal mode establishes a platform compatible with high-rate, scalable time-multiplexed photonic architectures.
\end{sciabstract}

\section*{Introduction}

Light offers a compelling platform for quantum information processing. Its carrier frequency in the hundreds of terahertz preserves quantum coherence under ambient conditions and enables ultrafast operation~\cite{Takeda2019}. In continuous-variable (CV) quantum information processing, quantum information is encoded in the quadrature amplitudes of optical fields and processed using Gaussian operations~\cite{Lloyd1999,Raussendorf2001,Alexander2018, Weedbrook2012,Braunstein2005, Asavanant2019, Larsen2019}. Because these operations can be implemented with mature optical components developed for optical communications, optical quantum computing offers a plausible route toward ultrafast quantum information processing~\cite{Kawasaki2025}.

However, universal, fault-tolerant quantum computation with Gaussian optical processors requires non-Gaussian resource states~\cite{Dakna1997,Gottesman2001,Hastrup2022, Konno2024, Larsen2025}. Two classes of such states are of particular practical interest: Schr\"{o}dinger cat states, which serve as codewords in cat-code bosonic encodings, and Gottesman--Kitaev--Preskill (GKP) qubits, which can be distilled from cat-like states by breeding-type protocols \cite{Vasconcelos2010}. In both cases, fault-tolerance arguments favor cat states with sufficiently large amplitudes~\cite{Hastrup2022,Takase2024,Lund2008,Guillaud2019, Cochrane1999, Mirrahimi2014, Leghtas2013} (the corresponding amplitude requirements for both encodings are detailed in the Supplementary Materials (S7)).

The generation rate of non-Gaussian resources is constrained by two largely independent factors: the success probability of the state-preparation protocol and the repetition rate limited by the temporal width of the optical wave packets. Most existing approaches rely on heralded preparation using linear-optical circuits closely related to Gaussian boson sampling (GBS)~\cite{Hamilton2017}, in which multiple squeezed states interfere in a beam-splitter network and selected output modes are measured with photon-number-resolving detectors~\cite{Su2019}. These methods are highly versatile and have recently succeeded in generating approximate GKP states exhibiting a $3\times3$ grid of negative regions in the Wigner function, heralded with a success probability of $\sim10^{-4}$ for the targeted photon-number pattern \cite{Larsen2025}. Reaching the still higher state quality required for fault tolerance is expected to be even more demanding: because the photon-number conditions become increasingly stringent, the corresponding success probability has been estimated to be on the order of $10^{-6}$ \cite{Takase2024}. At the same time, many demonstrations of complicated non-Gaussian states still operate with nanosecond-scale temporal modes. Shorter wave packets have been difficult to exploit because of limited source and detector bandwidth, detector timing jitter, imperfect temporal-mode matching, and dispersion, all of which increase the effective loss. As a result, the supply rate of non-Gaussian resources remains far below what is required for practical optical quantum computing.

Recent work suggests a route around both bottlenecks. On the protocol side, adaptive breeding can dramatically increase the success probability by applying Gaussian operations conditioned on measurement outcomes~\cite{Weigand2018}. This approach is particularly powerful when combined with generalized photon subtraction (GPS)~\cite{Takase2021} [Fig.~1(a)], for which theory predicts success probabilities for the GKP qubit on the order of $10^{-1}$ under realistic conditions~\cite{Takase2024}. On the hardware side, broadband squeezed-light sources~\cite{Kashiwazaki2021,Nehra2022}, fast photon-number-resolving detectors~\cite{Fukuda2024,Endo2025}, and ultrafast homodyne measurements are beginning to make high-repetition-rate operation feasible~\cite{Endo2023}.

Here we demonstrate multi-photon GPS in picosecond optical wave packets using a broadband optical parametric amplifier (OPA) with multi-terahertz bandwidth and a high-speed Ti--Au transition-edge sensor (TES). This platform enables the generation of Schr\"odinger cat states in $\sim$10-ps temporal modes, roughly two orders of magnitude shorter than the nanosecond-scale temporal modes employed in recent integrated GKP demonstrations~\cite{Larsen2025}. Using pulsed homodyne detection matched to the ultrashort wave packets, we reconstruct Wigner functions exhibiting up to four distinct negative regions without loss correction. The generated states reach an effective cat-state amplitude of $\alpha_{\rm eff} = 1.69$, approaching the amplitudes of practical relevance for fault-tolerant cat-code architectures~\cite{Lund2008,Guillaud2019,Hastrup2022} and for adaptive breeding toward GKP-state generation~\cite{Takase2024}, while preserving clear Wigner negativity (also see Supplementary Materials). These results establish the missing experimental capability for high-rate non-Gaussian state generation and provide a platform compatible with scalable optical quantum information processing.

\begin{figure}
\centering
\includegraphics[width=\textwidth]{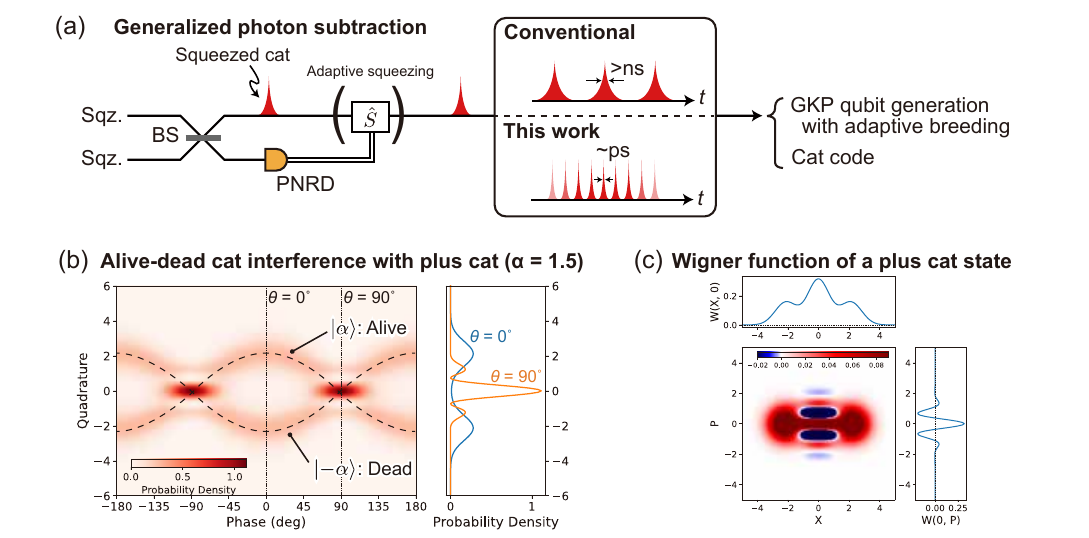}
\caption{\textbf{Generalized photon subtraction in picosecond wave packets for high-rate non-Gaussian-resource generation.} (a) Generalized photon subtraction (GPS) is implemented by interfering two squeezed states (Sqz.) on a beam splitter (BS) and heralding on photon-number-resolving detection (PNRD). The resulting non-Gaussian state (a squeezed cat state) can be further processed by Gaussian operations, including optional adaptive squeezing (\(\hat{S}\)), and serves as a resource for adaptive breeding toward GKP-state generation and cat-code quantum information processing. In this work, GPS is extended from conventional nanosecond-scale wave packets to picosecond temporal modes, thereby addressing the repetition-rate bottleneck in non-Gaussian-resource generation. (b) Phase-dependent quadrature distribution of a plus cat state, \(\ket{\alpha}+\ket{-\alpha}\), for \(\alpha=1.5\). At \(\theta=0^\circ\), the distribution shows two peaks centered at \(\pm\sqrt{2}\alpha\), whereas at \(\theta=90^\circ\) interference fringes appear between them. (c) Corresponding Wigner function of the plus cat state. The two displaced lobes and central interference fringes are characteristic of a coherent-state superposition, and the negative regions reflect its non-Gaussian quantum coherence.}\label{fig:intro}
\end{figure}

\section*{Results}

\subsection*{Cat states}

We begin by summarizing the structure of the Schr\"odinger cat states relevant to our experiment. In optics, a cat state is a quantum superposition of two coherent states,
\begin{align}
\ket{\mathrm{Cat}^{\pm}} \propto \ket{\alpha} \pm \ket{-\alpha},
\end{align}
where the components \(\ket{\alpha}\) and \(\ket{-\alpha}\) are often pictured as the ``alive'' and ``dead'' parts of the cat, respectively.

Figure~\ref{fig:intro}(b) shows the phase-dependent quadrature marginal distributions of a plus cat state, \(\ket{\mathrm{Cat}^{+}}\), with \(\alpha=1.5\). At \(\theta=0^\circ\), the marginal distribution exhibits two well-separated peaks corresponding to the two coherent-state components. In contrast, at \(\theta=90^\circ\), interference fringes appear between them, reflecting the quantum coherence of the superposition. This phase-dependent structure cannot arise from an incoherent mixture of \(\ket{\alpha}\) and \(\ket{-\alpha}\).

The corresponding Wigner function, shown in Fig.~\ref{fig:intro}(c), provides a phase-space representation of the same state. Two positive lobes correspond to the coherent-state components at \((\pm \sqrt{2}\alpha, 0)\), while interference fringes and negative regions emerge near the origin. These negative regions are a direct signature of non-Gaussian quantum coherence. Throughout this paper, we use the convention \(\hbar=1\), for which $\alpha = (x+ip)/\sqrt{2}$. Although the states generated in the present work are squeezed cat states, the same essential characteristics remain: the two-lobe quadrature structure, the interference fringes, and the negativity of the Wigner function. These features therefore serve as the primary benchmarks for evaluating the fidelity and effective cat-state amplitude of the generated non-Gaussian states.

Optical Schr\"odinger cat states have been generated by a variety of methods, including photon subtraction or addition from squeezed states \cite{Dakna1997, Ourjoumtsev2006, Neergaard-Nielsen2006, Gerrits2010, Ra2019, Endo2023, Chen2024}, breeding protocols \cite{Sychev2017}, and cavity-based approaches \cite{Hacker2019}. In the present work, we use GPS \cite{Takase2021}, in which two squeezed states interfere on a beam splitter and photon-number-resolving detection heralds a squeezed cat state [Fig.~\ref{fig:intro}(a)]. In particular, multi-photon-resolved GPS provides direct access to cat states with larger effective amplitudes, which is the regime favored by fault-tolerant bosonic encodings, making it a suitable route to generating large-amplitude non-Gaussian states in ultrashort temporal modes.

\begin{figure}
\centering
\includegraphics[width=\textwidth]{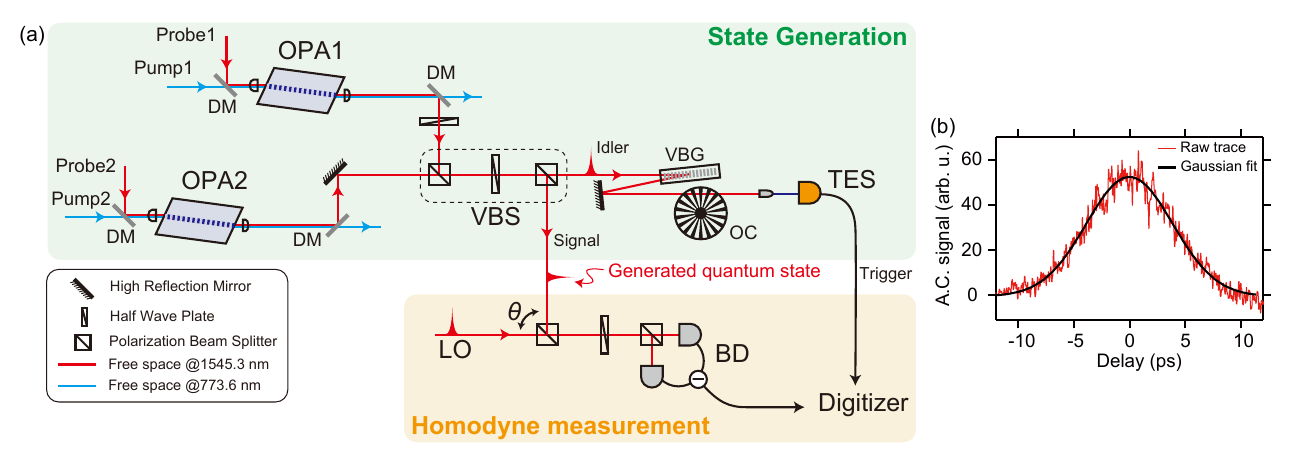}
\caption{(a) Experimental apparatus for quantum-state generation and measurement. Two optical parametric amplifiers (OPA1 and OPA2) generate squeezed-vacuum fields that are combined to prepare non-Gaussian quantum states via GPS. One output arm from the variable beam splitter (VBS) is spectrally filtered by a volume Bragg grating (VBG) and detected with a transition-edge sensor (TES)-based photon-number-resolving detector, providing the heralding trigger. The generated state in the other arm is analyzed by balanced homodyne detection using a balanced detector (BD) and a local oscillator (LO) with controllable phase \(\theta\), and the quadrature signals are recorded by a high-speed digitizer. DM, dichroic mirror; OC, optical chopper. (b) Autocorrelation (A.C.) trace of the LO pulse (red, measured data; black, Gaussian fit).}\label{fig:apparatus}
\end{figure}

\subsection*{Experimental apparatus}

As the GPS protocol itself has been described in detail elsewhere \cite{Takase2021}, here we focus on the experimental platform that enabled GPS at high repetition rates in picosecond temporal modes [Fig.~\ref{fig:apparatus}(a)]. Two independently phase-controlled squeezed-vacuum states were generated in broadband type-0 periodically poled lithium niobate waveguide OPAs (OPA1 and OPA2; see Methods), pumped by 772.66-nm pulses with a repetition rate of 5 MHz and a duration of \(\sim 10\) ps. Degenerate parametric down-conversion produced squeezed light at 1545.32 nm. Auxiliary probe beams were used to stabilize both the squeezing phases and the interferometric phase of the GPS circuit.

The two squeezed modes were interfered on a variable beam splitter (VBS), which consisted of two polarizing beam splitters and a half-wave plate. One output, after spectral filtering with a volume Bragg grating (VBG) and removal of the probe light with an optical chopper (OC), was sent to a high-speed Ti--Au TES for photon-number-resolved heralding (see Methods). The other output was analyzed by pulsed homodyne detection using a local oscillator (LO) centered at 1545.32 nm. The LO pulses were shaped to match the ultrashort temporal mode of the generated state, and their autocorrelation trace [Fig.~\ref{fig:apparatus}(b)] yielded a full width at half maximum of 10.0 ps. Because homodyne detection projects onto the spatiotemporal mode of the LO, this width defines the temporal mode of the reconstructed quantum states. Temporal and spatial mode overlaps exceeded 90\% and 98\%, respectively.

The TES and homodyne outputs were acquired simultaneously with a high-speed digitizer. An onboard FPGA analyzed the TES waveform in real time and, for events in which two or more photons were detected, recorded the TES pulse height together with the corresponding homodyne data. This enabled photon-number-resolved analysis of the generated states. For the measurements reported here, the parametric gains of OPA1 and OPA2 were 6.5 and 7.4 dB, respectively, the beam-splitter reflectivity was set to 35\%, and the interference phase was fixed at \(90^\circ\).

Quantum-state tomography was performed from more than \(10^8\) homodyne samples acquired over 12 quadrature settings. Density matrices were reconstructed by maximum-likelihood estimation without loss correction \cite{Lvovsky2004}.

\subsection*{Generated cat states}

\begin{figure}
\centering
\includegraphics[width=0.75\textwidth]{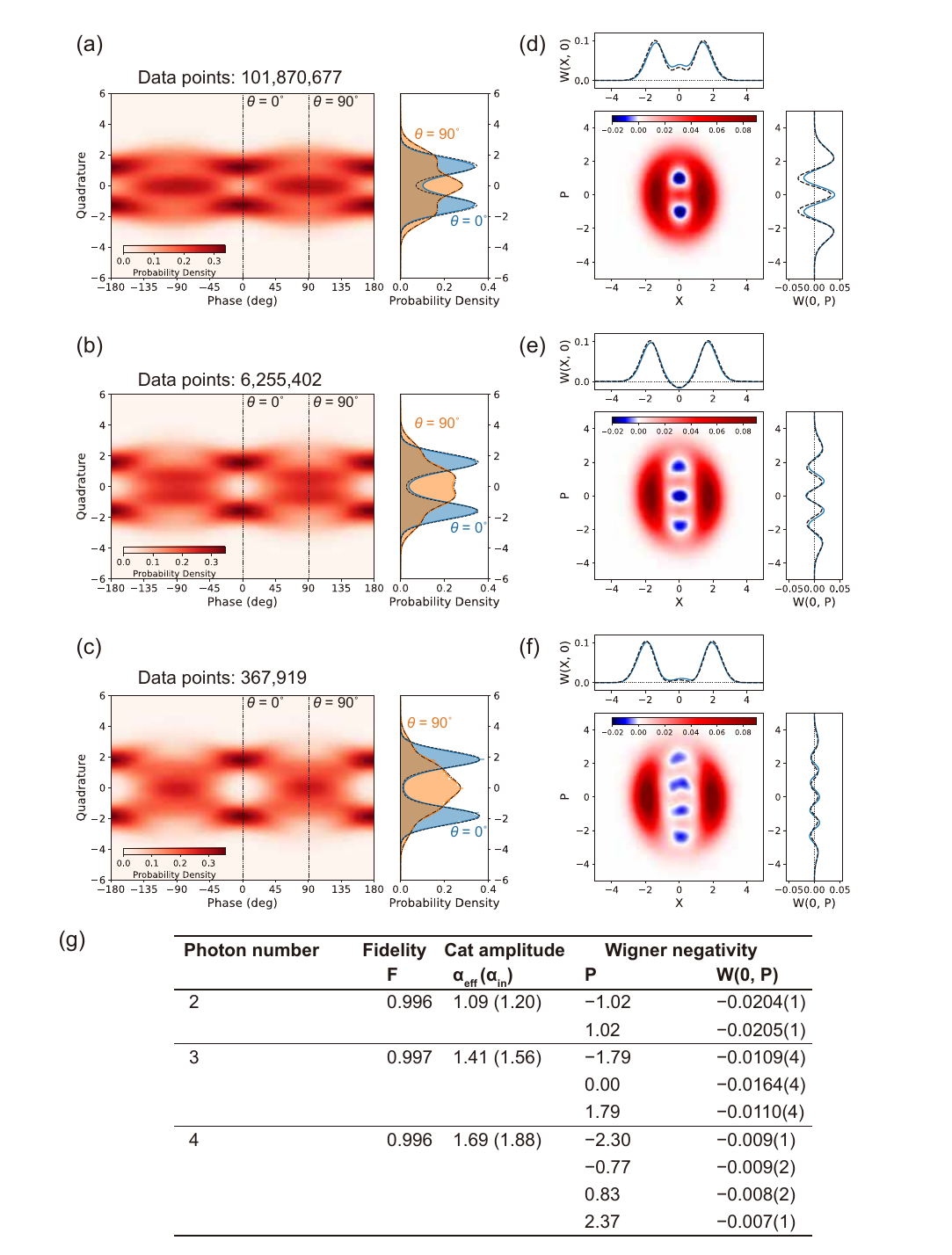}
\caption{(a--c) Phase-dependent quadrature marginal distributions of the states conditioned on photon-number detection events with \(n=2\), 3, and 4. Heat maps show the angle-resolved marginals. The right-hand panels compare cross sections at \(\theta=0^\circ\) and \(90^\circ\) with the corresponding experimental histograms (bars), reconstructed marginals (solid lines), and simulations (dashed and dash-dotted lines). The number of quadrature samples used for the quantum-state tomography is indicated at the top of each heat map. (d--f) Reconstructed Wigner functions for \(n=2\), 3, and 4, together with cross sections along the \(X\) and \(P\) axes (solid lines). The dashed lines show the corresponding simulations. (g) Fidelity to the numerically simulated state, effective cat-state amplitude \(\alpha_{\mathrm{eff}}\) (with \(\alpha_{\mathrm{in}}\) in parentheses), and measured Wigner minimum \(W(0,P)\) at the indicated phase-space coordinate \(P\). Parameters are extracted from homodyne tomography; uncertainties are shown in parentheses.}\label{fig:results}
\end{figure}

Figure~\ref{fig:results}(a)--(c) shows the phase-dependent quadrature marginal distributions of the states heralded by detection of \(n=2\), 3, and 4 photons. In the right-hand panels, the measured histograms at \(\theta=0^\circ\) and \(90^\circ\) (blue and orange bars, respectively) are compared with the corresponding marginal distributions obtained from the reconstructed states (blue and orange solid lines) and with simulations based on independently calibrated squeezing and loss (dashed and dash-dotted lines). The close agreement among measurement, reconstruction, and simulation indicates that both the state generation and the homodyne measurement are accurately described. As the herald photon number \(n\) increases, the separation between the two peaks at \(\theta=0^\circ\) becomes larger, while the distribution at \(\theta=90^\circ\) develops more pronounced oscillatory structure. Both trends are consistent with the generation of cat states with larger effective amplitudes. The reconstructed states also agree well with the simulations; as summarized in Fig.~\ref{fig:results}(g), the fidelities,
\begin{align}
F = \mathrm{Tr}\!\left[\left(\sqrt{\rho_{\mathrm{exp}}}\,\rho_{\mathrm{sim}}\,\sqrt{\rho_{\mathrm{exp}}}\right)^{1/2}\right],
\end{align}
where \(\rho_{\mathrm{sim}}\) and \(\rho_{\mathrm{exp}}\) are the density matrices obtained by simulation and experiment, respectively, exceed 0.99 for \(n=2\), 3, and 4.

To quantify this growth in cat-state amplitude, we numerically applied an inverse squeezing operation to each reconstructed state until the widths of the two peaks at \(\theta=0^\circ\) matched the vacuum width (see Supplementary Materials). The effective cat-state amplitude was then extracted from the resulting peak separation. The results are summarized in Fig.~\ref{fig:results}(g). Without loss correction, we obtain \(\alpha_{\mathrm{eff}}=1.09\), 1.41, and 1.69 for \(n=2\), 3, and 4, respectively. Thus, photon-number-resolved GPS provides direct access to progressively larger cat states, reaching \(\alpha_{\mathrm{eff}}=1.69\) for \(n=4\). This value lies in the regime of practical relevance for cat-code architectures and approaches that are required by adaptive breeding for GKP-state generation, as discussed in detail below. Correcting only for the 80\% homodyne efficiency gives \(\alpha_{\mathrm{in}}=1.20\), 1.56, and 1.88.

Note that this amplitude definition differs from that commonly used in previous optical cat-state experiments \cite{Sychev2017}. In earlier works, the reconstructed state is first corrected for all known losses, and the cat-state amplitude is then inferred from the squeezed cat state that maximizes the fidelity to the corrected state. Because such a procedure deconvolves loss, it generally yields larger amplitudes. By contrast, the \(\alpha_{\mathrm{eff}}\) values reported here are extracted directly from the uncorrected quadrature marginals and therefore more closely reflect the experimentally observed states. The partially corrected values \(\alpha_{\mathrm{in}}\), which account only for the homodyne efficiency, are given as a reference. Direct numerical comparison between the present amplitudes and literature values obtained from fully loss-corrected, fidelity-based estimates is therefore not straightforward.

The corresponding Wigner functions reconstructed without loss correction are shown in Fig.~\ref{fig:results}(d)--(f). All three states exhibit clear negativity, providing direct evidence of non-Gaussianity. Moreover, as \(n\) increases, the number of distinct negative regions grows from two to four, consistent with the increase in cat-state amplitude. The minima of \(W(0,P)\) are summarized in Fig.~\ref{fig:results}(g). Error bars on the Wigner minima were estimated by bootstrap resampling (see Supplementary Materials). Because the event rate decreases with increasing photon number, insufficient statistics were available for reliable tomography of the \(n=5\) state; only preliminary data are therefore provided in the Supplementary Materials.

\section*{Discussion}

Our results establish multi-photon GPS in picosecond optical wave packets, addressing a central bottleneck in optical CV quantum information processing: the generation of non-Gaussian resources at rates compatible with ultrafast optical hardware. Previous demonstrations of strongly non-Gaussian optical states have largely remained in nanosecond temporal modes. By contrast, the present platform maintains multi-photon heralded state preparation and homodyne characterization in ${\sim}10$-ps wave packets while preserving clear Wigner negativity without loss correction. The integration of broadband squeezing, a high-speed Ti--Au TES, and picosecond-mode pulsed homodyne detection constitutes the key enabling capability of this work.

The effective amplitude $\alpha_{\mathrm{eff}} = 1.69$, obtained without loss correction, brings our states close to the amplitude range of practical relevance for cat-code architectures, in which the bit-flip rate is suppressed exponentially as $\Gamma_{\mathrm{bit}} \propto e^{-2\alpha^2}$ while the phase-flip rate grows only linearly as $\Gamma_{\mathrm{phase}} \propto \alpha^2$~\cite{Lund2008,Guillaud2019}, giving rise to the biased-noise structure that underlies hardware-efficient fault tolerance. This inverts the longstanding intuition that small-amplitude ``kittens'' are preferable for their robustness against loss: in bosonic quantum error correction, cat states function as codewords whose code distance grows with $\alpha$ \cite{Bergmann2016, Li2017, Grimsmo2020}, and our platform supplies such states directly in picosecond temporal modes.

A second implication concerns GKP-state generation by adaptive breeding~\cite{Takase2024,Weigand2018}, where these works removed the post-selection requirement of the original proposal~\cite{Vasconcelos2010}. In an $m$-step tree-type protocol, each step contracts the neighboring-peak spacing by $\sqrt{2}$, so an $m = 2$ protocol contracts it by a factor of $2$. Reaching the symmetric qunaught grid state---a fully symmetric oscillator grid state of period $\sqrt{2\pi}$ (in our $\hbar = 1$ convention) that serves as an elementary resource for higher-level GKP processing~\cite{Walshe2020}---thus requires an initial cat-state peak separation of $2\sqrt{2\pi}$, corresponding to an amplitude $\alpha \approx \sqrt{\pi} \approx 1.77$ (see Supplementary Materials~S7). Our $\alpha_{\mathrm{eff}} = 1.69$ lies within ${\sim}5\%$ of this target while preserving clear Wigner negativity, supplying the missing ingredient---high-fidelity multi-photon GPS at large $\alpha$ in short temporal modes---for breeding protocols whose success probability has been predicted to reach $10^{-1}$~\cite{Takase2024}.

We note again that approximate GKP states have recently been realized through a GBS-type (non-adaptive) preparation in nanosecond temporal modes~\cite{Larsen2025}. The adaptive breeding route considered here is a separate strategy, in which measurement outcomes condition subsequent Gaussian operations on multi-photon GPS resources, and is theoretically predicted to offer substantially higher success probabilities under realistic conditions~\cite{Takase2024}.

The repetition rate in the current implementation is limited mainly by  the timing jitter of the TES (${\sim}50$~ns) and the bandwidth of the homodyne receiver (${\sim}25$~MHz)---technical rather than conceptual constraints. Faster photon-number-resolving detectors such as SNSPD-based implementations~\cite{Cheng2022, Sidorova2025,Endo2025}, together with OPA-assisted pre-amplified homodyne detection~\cite{Inoue2023,Kalash2023}, should extend operation toward the hundreds-of-megahertz or even gigahertz regime, directly increasing the supply rate of cat and GKP resources.

More broadly, this work moves non-Gaussian state generation toward the operating conditions required by scalable optical quantum processors, establishing a practical route to high-rate non-Gaussian resource generation for adaptive breeding, bosonic encoding, and time-multiplexed optical quantum information processing.

\section*{Methods}
\subsection*{Details of the experiment}
Here, we describe the details of the experiment. The detailed phase-lock method is described in our previous paper \cite{Endo2023}.

\subsubsection*{Broadband squeezed-light sources}

Two squeezed-vacuum states were generated with two type-0 periodically poled lithium niobate (PPLN) waveguide optical parametric amplifiers (OPA1 and OPA2), pumped by pulses at a center wavelength of 772.66 nm with a repetition rate of 5 MHz and a pulse duration of approximately 10 ps. The pulse shape was controlled by a home-made pulse shaper, and degenerate parametric down-conversion produced squeezed light at 1545.32 nm. Each waveguide was \SI{10}{mm} long, with a core of width approximately \SI{9}{\micro m} and thickness approximately \SI{8}{\micro m}; the core consisted of ZnO-doped LN directly bonded to a lithium tantalate substrate. The periodic poling structure for quasi-phase matching (QPM), with a poling cycle of approximately \SI{18}{\micro m}, was fabricated using an electrical poling method \cite{Umeki2010}, and the waveguide structure itself was fabricated using a mechanical dicing saw; further fabrication details are given in Ref.~\cite{Kashiwazaki2021}. Both end faces of the waveguide were coated with anti-reflective coatings for 1545.32-nm and 772.66-nm light, and the waveguide temperature was controlled to maintain the QPM condition. The parametric gains of the two OPAs were independently adjustable from 0 to 8 dB.

Co-propagating probe beams were injected into both OPAs to monitor and stabilize the squeezing phases. Owing to phase-sensitive amplification in the waveguides, the probe beams carried the phase information of the squeezed fields and were used to stabilize the downstream interferometer.

\subsubsection*{Ti-Au TES}
The TES device was fabricated at the superconducting quantum circuit fabrication facility (Qufab) at AIST. The superconducting TES film comprises bilayers of titanium and gold, deposited by DC magnetron sputtering to thicknesses of \SI{20}{nm} and \SI{10}{nm}, respectively. To enhance detection efficiency, an optical cavity structure was adopted, in which the TES films are embedded between a high-reflection mirror and anti-reflection coatings. The TES is patterned into a square with dimensions of \SI{5}{\micro m} $\times$ \SI{5}{\micro m}. Superconducting niobium leads were fabricated on two opposite sides of the TES, and the resulting resistance changes are read via these leads using a SQUID-based current amplifier.

To deliver photons, a high-NA optical fiber (UHNA7, Nufern) with a mode field diameter (MFD) of \SI{3.2}{\micro m} was coupled to the TES device. The fiber-coupled TES module was installed in an adiabatic demagnetization refrigerator and cooled to below \SI{280}{mK}. Inside the refrigerator, the high-NA optical fiber was spliced to an SMF-28 fiber (Corning) to convert the MFD to \SI{10}{\micro m}, with an MFD conversion loss of less than \SI{0.2}{dB}.

The critical temperature and normal resistance of the TES are \SI{308.6}{mK} and \SI{2.653}{\ohm}, respectively. The TES exhibits an energy resolution of $\Delta E=\SI{0.176}{eV}$, corresponding to an $E/\Delta E$ ratio of 4.54 for a single-photon energy of \SI{0.8}{eV} at a wavelength of \SI{1.5}{\micro m}, thereby demonstrating sufficient capability to resolve individual photon states. The system detection efficiency of the TES was carefully calibrated using an evaluation system traceable to the National Metrology Institute of Japan (NMIJ) at AIST, yielding an efficiency of 89.2\% at a wavelength of \SI{1.55}{\micro m}. The decay time constant of the observed signals was \SI{107}{ns}, enabling a repetition frequency of up to \SI{5}{MHz}.

\subsubsection*{Pulsed homodyne detection and temporal-mode matching}

The signal arm was analyzed by a home-made balanced homodyne detector using a pulsed LO light centered at 1545.32 nm. The efficiency of the photodiodes is 98\%, and the diode's diameter is \SI{100}{\micro m}. The temporal mode of the local oscillator was shaped with a wave shaper (WS-01000A, Coherent) to match the generated ultrashort quantum state.

Temporal mode matching between the signal and local oscillator was about 90\%. Spatial mode matching was evaluated independently with a continuous-wave laser and exceeded 98\%. Taking these contributions together with the photodiode quantum efficiency and the electronic loss into account, the overall effective homodyne detection efficiency was estimated to be 80\%.

\subsubsection*{Data acquisition}

The output signals of the TES and the homodyne detector were digitized simultaneously by the digitizer (M5200A, Keysight). For each heralding condition, homodyne data were acquired at 12 quadrature settings. Each setting contained approximately \(8\times 10^6\) samples, giving a total data set exceeding \(10^8\) points. These large data sets were required to reconstruct states exhibiting multiple Wigner-function negativities with sufficient statistical confidence.

\subsubsection*{Quantum-state reconstruction}

Density matrices were reconstructed from the homodyne data by maximum-likelihood quantum-state tomography \cite{Lvovsky2004}. Unless otherwise stated, no loss correction was applied. Wigner functions were calculated directly from the reconstructed density matrices.

To estimate the effective cat-state amplitude, an inverse squeezing operation was numerically applied to each reconstructed state until the widths of the two peaks in the \(\theta=0^\circ\) quadrature marginal matched that of the vacuum state. The amplitude was then extracted from the corresponding peak separation. The values denoted \(\alpha_{\mathrm{eff}}\) were obtained without loss correction. For reference, \(\alpha_{\mathrm{in}}\) was obtained by correcting only for the measured homodyne efficiency of 80\%.

Error bars on the Wigner minima were estimated by bootstrap resampling. We used \(N=100\) resamples for the \(n=2\) and \(n=3\) data and \(N=500\) resamples for the \(n=4\) data.

\subsection*{Simulation}
We performed numerical simulations using Strawberry Fields \cite{Killoran2019, Bromley2020}, and the full simulation code is publicly available at Zenodo (see Data and materials availability). The model implements GPS and the subsequent measurement process. Two squeezed-vacuum states---with squeezing parameters $(r_\mathrm{OPA1}, r_\mathrm{OPA2}) = (0.748, 0.852)$, which correspond to the parametric gains of  of 6.5 and 7.4 dB, and optical losses $l_\mathrm{OPA1,2} = 0.07$---are interfered on a beam splitter with reflectivity $R=0.35$ and an interference phase of $\theta = 90^\circ$. One output port is subjected to photon-number-resolving detection with efficiency $\eta_{\mathrm{Idler}} = 0.66$ (including the fiber coupling, spectrum filtering, and $\eta_\mathrm{TES}$), while the other is measured by homodyne detection with efficiency $\eta_{\mathrm{HD}}=0.80$ and the propagation loss of $0.063$. Simulated Wigner functions are provided in the Supplementary Materials.

\section*{Acknowledgments}
This work was partly supported by Japan Science and Technology (JST) Agency (Moonshot R\&D, Grant No. JPMJMS2064, and PRESTO, Grant No. JPMJPR2254),  Council for Science, Technology and Innovation (CSTI), Cross-ministerial Strategic Innovation Promotion Program (SIP), ``Promoting Application of Advanced Quantum Technologies to Social Challenges'' (Project management agency: QST), the UTokyo Foundation, and donations from Nichia Corporation.
P.M. and P.L. acknowledge the European Union's HORIZON Research and Innovation Actions under Grant Agreement no. 101080173 (CLUSTEC).
P.M. acknowledges the project CZ.02.01.01\/00\/22\_008\/0004649 (QUEENTEC) of EU and the Czech Ministry of Education, Youth and Sport, and 25-17472S of the Czech Science Foundation.
R.F. acknowledges the project No. 21-13265X of the Czech Science Foundation.
P.L. acknowledges the BMFTR (former BMBF) in Germany for their support via the projects QR.N and PhotonQ.
The authors used ChatGPT (OpenAI) and Claude (Anthropic) to assist with generating analysis code and for minor grammatical editing of the manuscript. All scientific content, interpretations, and conclusions were developed and verified by the authors.

\section*{Author contributions}
M.E. conceived the project and led the experiment. M.E. built the optical and electrical setup with support from T.N., T.S., and K. Takahashi. M.E. coded the control program and the FPGA program for data acquisition with support from T.N.. M.E., P.M., and R.F. performed simulations and theory evaluation, and analyzed the data. T.K., A.I., and T.U. provided the OPA used in the experiment. S.T. and D.F. provided the TES used in the experiment. M.E. wrote the manuscript with P.L., P.M., R.F., K. Takase, W.A., A.F., and all the co-authors.

\section*{Competing interests}
The authors declare no competing interests.

\section*{Data and materials availability}
All data needed to evaluate the conclusions in the paper are present in the paper and/or the Supplementary Materials. The simulation/analysis code and raw data are available at Zenodo (DOI: 10.5281/zenodo.20640586).

\bibliography{scibib}

\bibliographystyle{Science}

\section*{Supplementary Materials}

\setcounter{figure}{0}
\setcounter{table}{0}
\setcounter{equation}{0}
\renewcommand{\thefigure}{S\arabic{figure}}
\renewcommand{\thetable}{S\arabic{table}}
\renewcommand{\theequation}{S\arabic{equation}}

\makeatletter
\renewcommand\citeform[1]{S#1}
\renewcommand\@biblabel[1]{S#1.}
\makeatother

\subsection*{S1. Quadrature data conditioned on photon-number measurement}

For each photon number identified by the TES, we measured the corresponding quadrature amplitudes of the generated state using pulsed balanced homodyne detection. As described in the main text, the quadrature data were acquired at 12 phase settings. We also show the corresponding data for \(n=5\).

\begin{figure}[p]
\centering
\includegraphics[width=\textwidth]{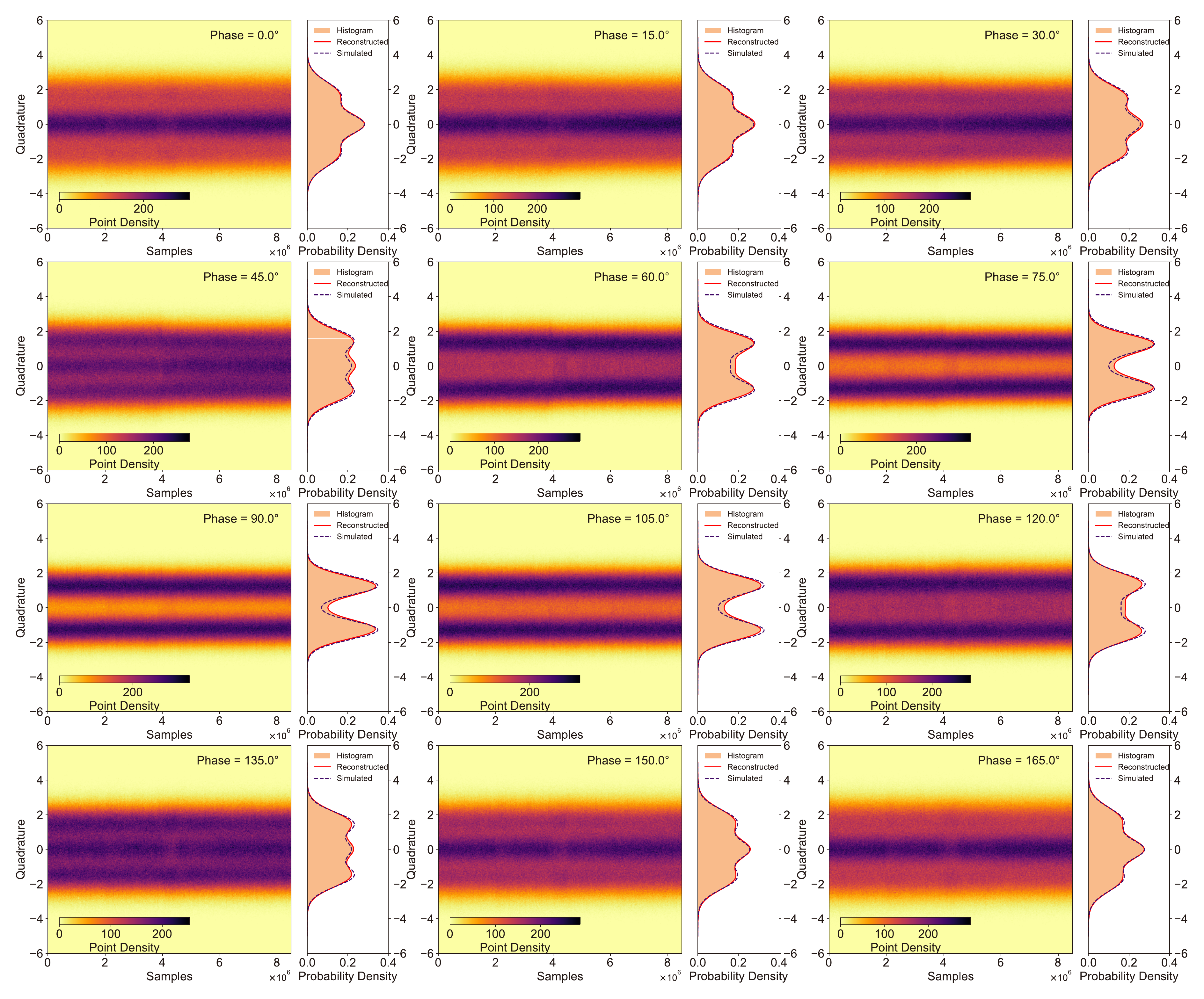}
\caption{Single-shot quadrature data for the \(n=2\) conditioned state. The scatter plot shows individual homodyne outcomes acquired at a fixed local-oscillator (LO) phase indicated in the panel. The histogram is constructed from the same data. The solid curve and dashed curve show the marginal distributions calculated from the reconstructed density matrix and from the numerical simulation, respectively.}
\label{fig:S_2photons}
\end{figure}

\begin{figure}[p]
\centering
\includegraphics[width=\textwidth]{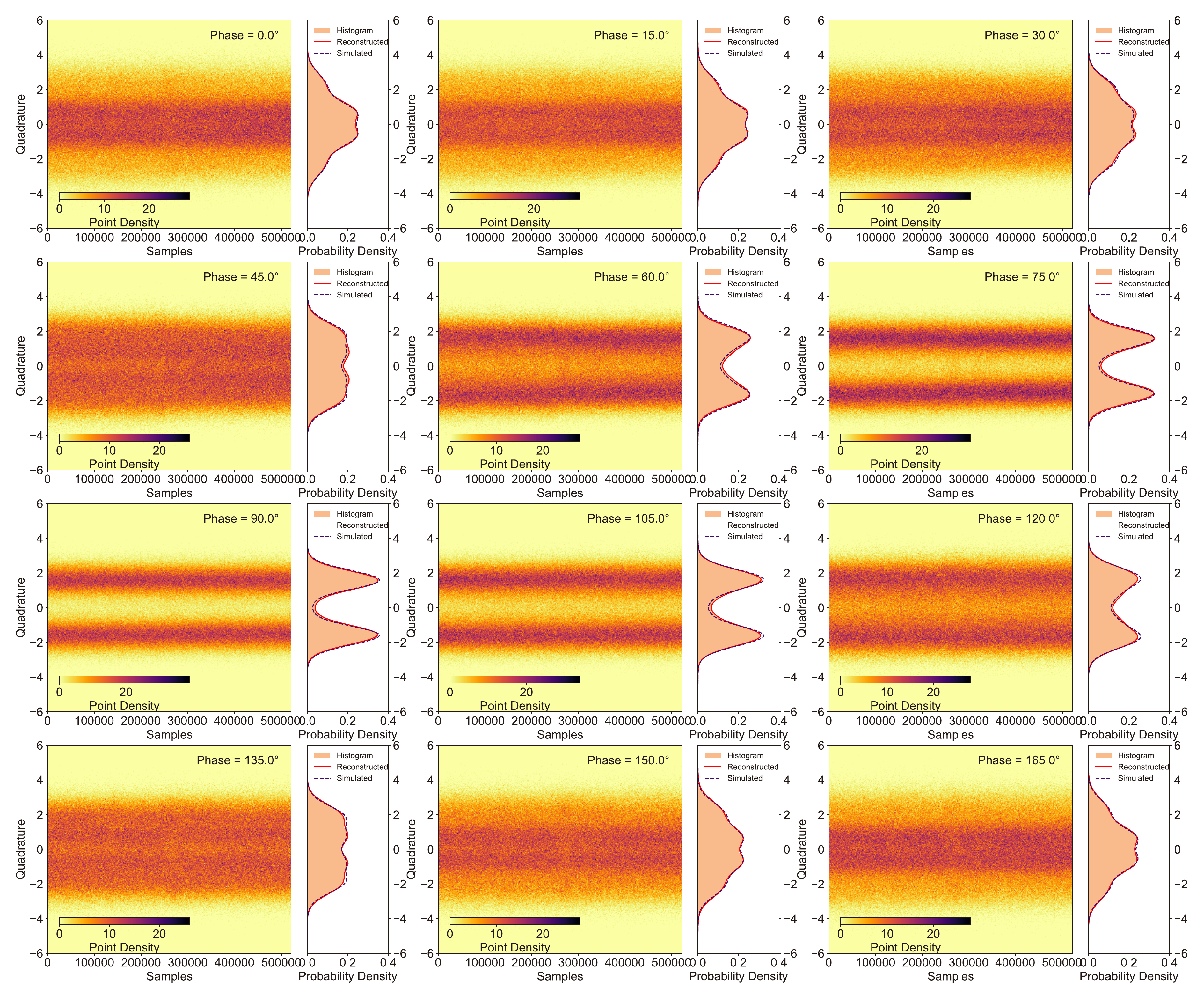}
\caption{Single-shot quadrature data for the \(n=3\) conditioned state. The plotting convention is the same as in Fig.~\ref{fig:S_2photons}.}
\label{fig:S_3photons}
\end{figure}

\begin{figure}[p]
\centering
\includegraphics[width=\textwidth]{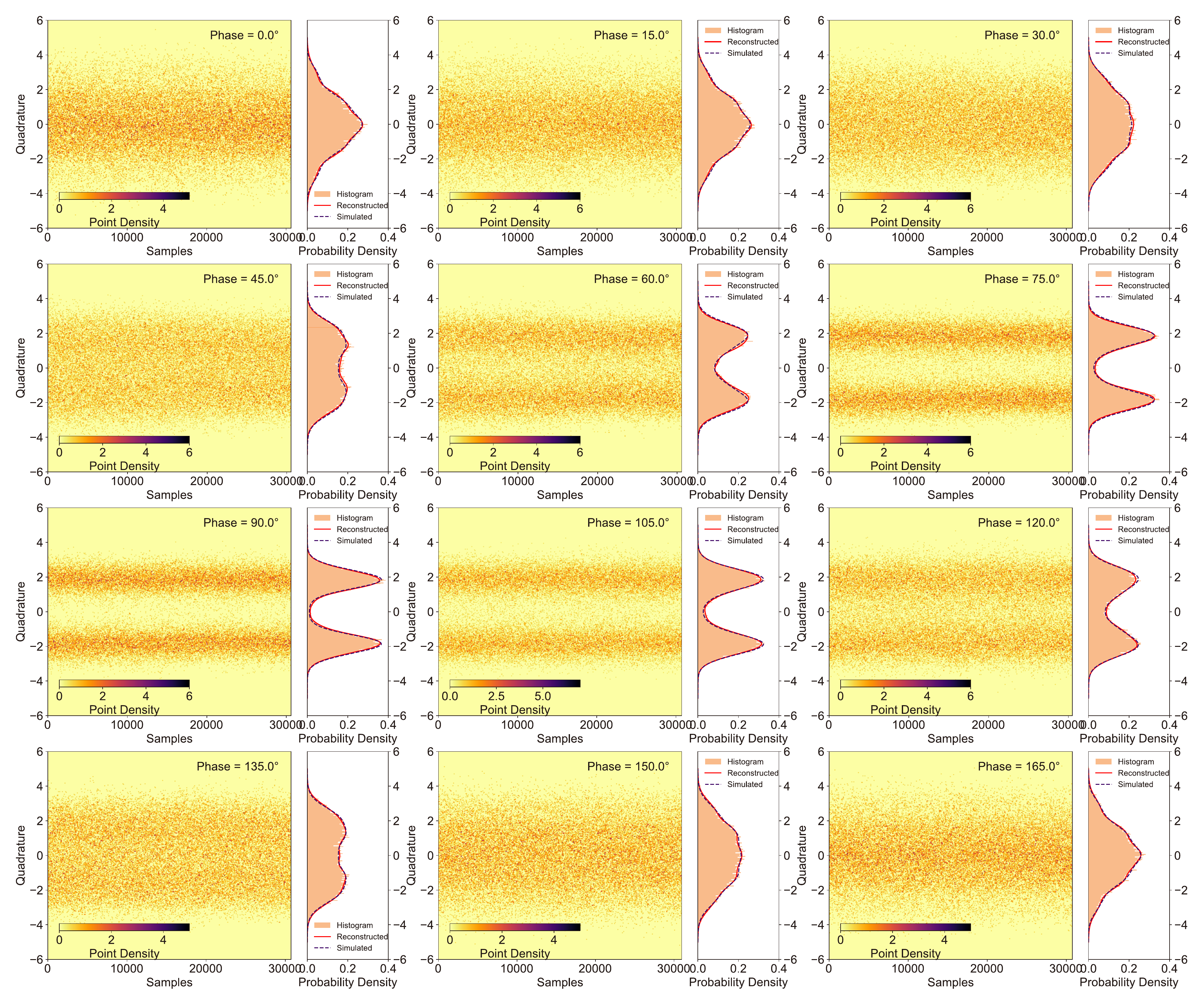}
\caption{Single-shot quadrature data for the \(n=4\) conditioned state. The plotting convention is the same as in Fig.~\ref{fig:S_2photons}.}
\label{fig:S_4photons}
\end{figure}

\begin{figure}[p]
\centering
\includegraphics[width=\textwidth]{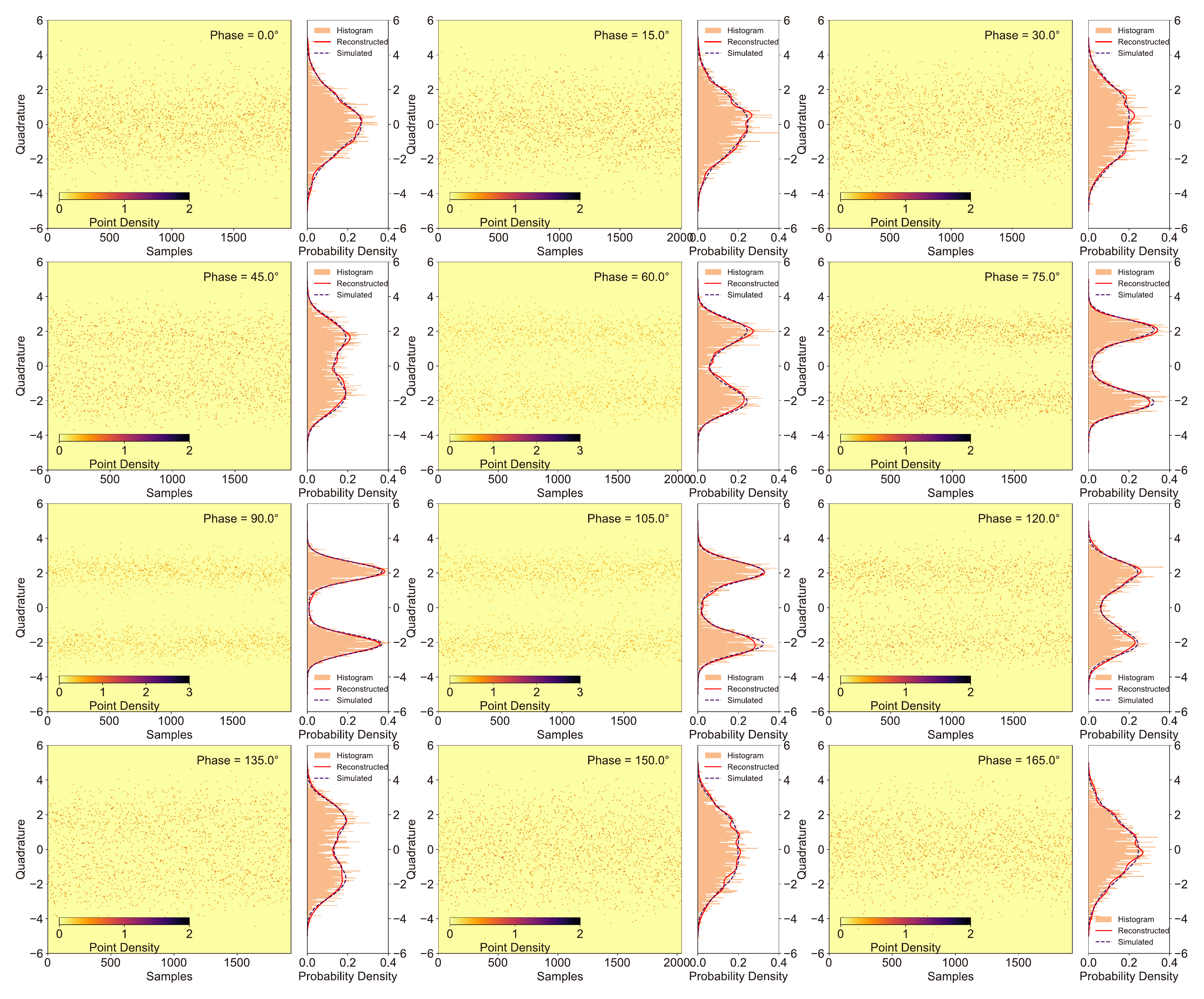}
\caption{Single-shot quadrature data for the \(n=5\) conditioned state. The plotting convention is the same as in Fig.~\ref{fig:S_2photons}. These data are shown for qualitative reference only.}
\label{fig:S_5photons}
\end{figure}

\paragraph*{Quadrature calibration.}
The scatter plots in Figs.~\ref{fig:S_2photons}--\ref{fig:S_5photons} show individual quadrature outcomes recorded at fixed LO phase. The vertical axis denotes the calibrated quadrature amplitude, and the horizontal axis denotes the sample index. Each point corresponds to one heralded event.

The output of the homodyne detector is a voltage signal. To convert this voltage into a dimensionless quadrature amplitude, shot-noise calibration was performed under the same timing conditions as the signal measurement.

\paragraph*{Shot-noise calibration.}
Shot noise was measured by blocking the signal path while keeping the LO active. Throughout this work, we use the convention \(\hbar=1\), such that the vacuum quadrature distribution is Gaussian with zero mean and variance \(\mathrm{Var}(x)=\mathrm{Var}(p)=1/2\). From the shot-noise voltage histogram, we determined a constant offset and a scaling factor so that the calibrated distribution had mean zero and variance \(1/2\). The same offset and scaling factor were applied to all datasets.

\paragraph*{Long-term stability.}
The measurements were conducted over several tens of hours, including the ADR recharging time. To ensure stability over this time scale, active stabilization was implemented for the LO power and the interferometric phases. Residual drifts were sufficiently small to be neglected within the statistical uncertainty, and a single calibration was used consistently throughout the measurements.

\paragraph*{Comparison with tomography and simulation.}
For each phase setting, the histogram (orange bars) shown in each panel is constructed directly from the single-shot quadrature outcomes. The red solid curve represents the corresponding marginal distribution calculated from the density matrix reconstructed by homodyne tomography. The dashed curve represents the marginal distribution obtained from the numerical simulation.

\FloatBarrier
\subsection*{S2. Density matrices in the Fock basis}

For each conditioned state, we reconstructed the density matrix from the homodyne quadrature data by maximum-likelihood estimation. The real and imaginary parts of the density matrix in the photon-number (Fock) basis are shown for all measured states.

The Hilbert-space truncation used in the reconstruction is identical to that employed in the main text. Convergence with respect to the cutoff dimension was verified by confirming negligible population near the boundary of the truncated Hilbert space.

\begin{figure}[htbp]
\centering
\includegraphics[width=0.75\textwidth]{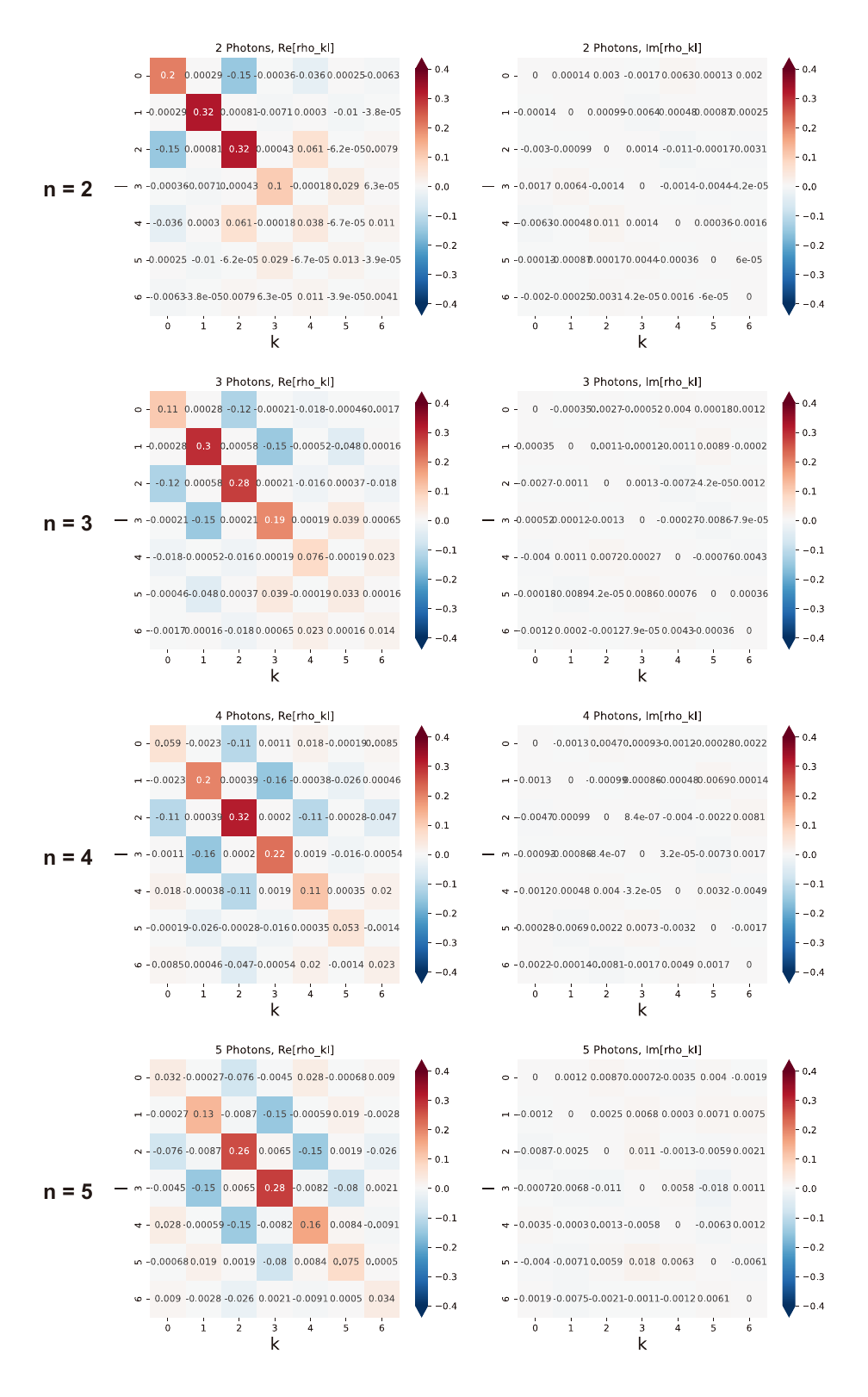}
\caption{Real and imaginary parts of the reconstructed density matrices in the Fock basis ($\rho_{kl}$) for each photon-number-conditioned state.}
\label{fig:S_density}
\end{figure}

\FloatBarrier
\subsection*{S3. TES pulse-height histogram and photon-number discrimination}

\begin{figure}[htbp]
\centering
\includegraphics[width=0.4\textwidth]{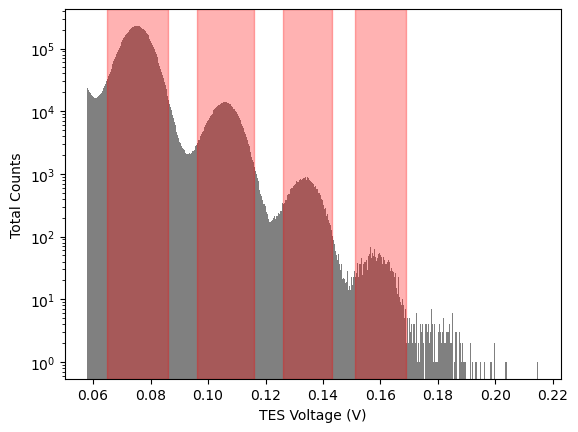}
\caption{TES pulse-height histogram used for photon-number discrimination. Because the acquisition trigger was set to \(n\ge2\), the lowest-voltage peak corresponds to \(n=2\) events. The red-shaded regions indicate the voltage windows used to extract events for each \(n\).}
\label{fig:SI_TES_hist}
\end{figure}

Figure~\ref{fig:SI_TES_hist} shows the pulse-height histogram of the transition-edge sensor (TES) used as the photon-number-resolving detector. The horizontal axis represents the TES pulse height, and the vertical axis represents the number of events.

In this experiment, the acquisition trigger was configured to record only events with two or more detected photons. As a result, the lowest-voltage peak in the histogram corresponds to \(n=2\) events.

In the raw dataset, the TES pulse height and the corresponding homodyne outcome were stored as paired values for each trigger event. Photon-number-conditioned subsets were then extracted offline by applying discrimination thresholds to the TES pulse height.

The red-shaded regions in Fig.~\ref{fig:SI_TES_hist} indicate the voltage windows used to assign each photon-number outcome.

\FloatBarrier
\subsection*{S4. Error-bar estimation}

\begin{figure}[htbp]
\centering
\includegraphics[width=\textwidth]{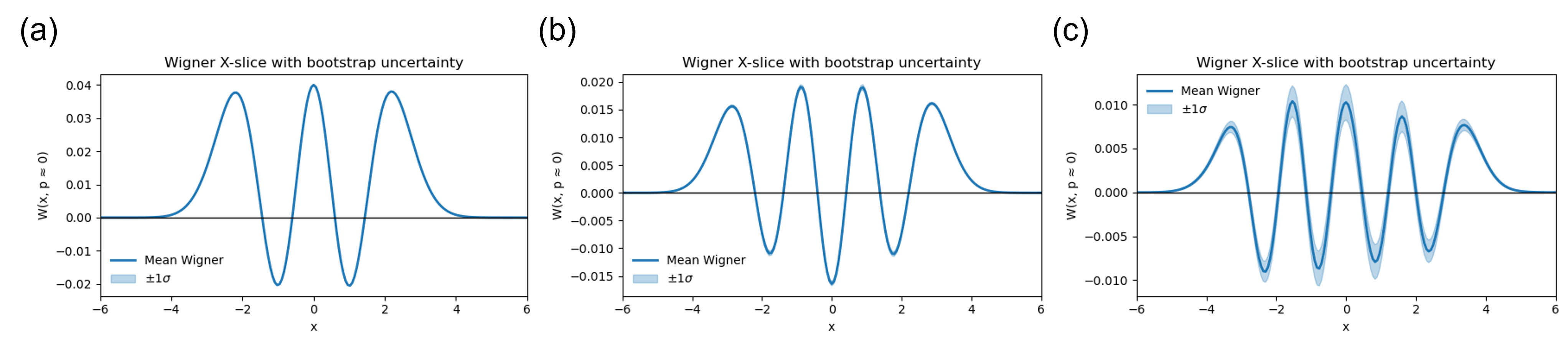}
\caption{Cross sections of the reconstructed Wigner functions used to evaluate the Wigner minima reported in the main text. Solid curves show the mean reconstructed Wigner functions, and the shaded regions indicate the \(\pm 1\sigma\) confidence intervals obtained from bootstrap resampling. For the \(n=2\) and \(n=3\) states, the bootstrap uncertainties are smaller than the line width and are therefore not visible.}
\label{fig:SI_Wigner_err}
\end{figure}

All error bars reported in this work were estimated by bootstrap resampling. The uncertainties of the Wigner-function minima are given in the main text. For completeness, Fig.~\ref{fig:SI_Wigner_err} shows the corresponding cross sections of the reconstructed Wigner functions together with the bootstrap confidence intervals.

For the bootstrap procedure, we used 100 resamples for the \(n=2\) and \(n=3\) datasets and 500 resamples for the \(n=4\) and \(n=5\) datasets.

\FloatBarrier
\subsection*{S5. Numerical simulation and comparison with experiment}

Numerical simulations were performed with Strawberry Fields \cite{Killoran2019, Bromley2020}.

The fidelities between the experimentally reconstructed density matrices and the simulated states exceed 0.99 for all states reported in the main text. This strong agreement confirms that the independently characterized experimental parameters capture the dominant physical processes governing both the state generation and the measurement.

\begin{figure}[htbp]
\centering
\includegraphics[width=0.7\textwidth]{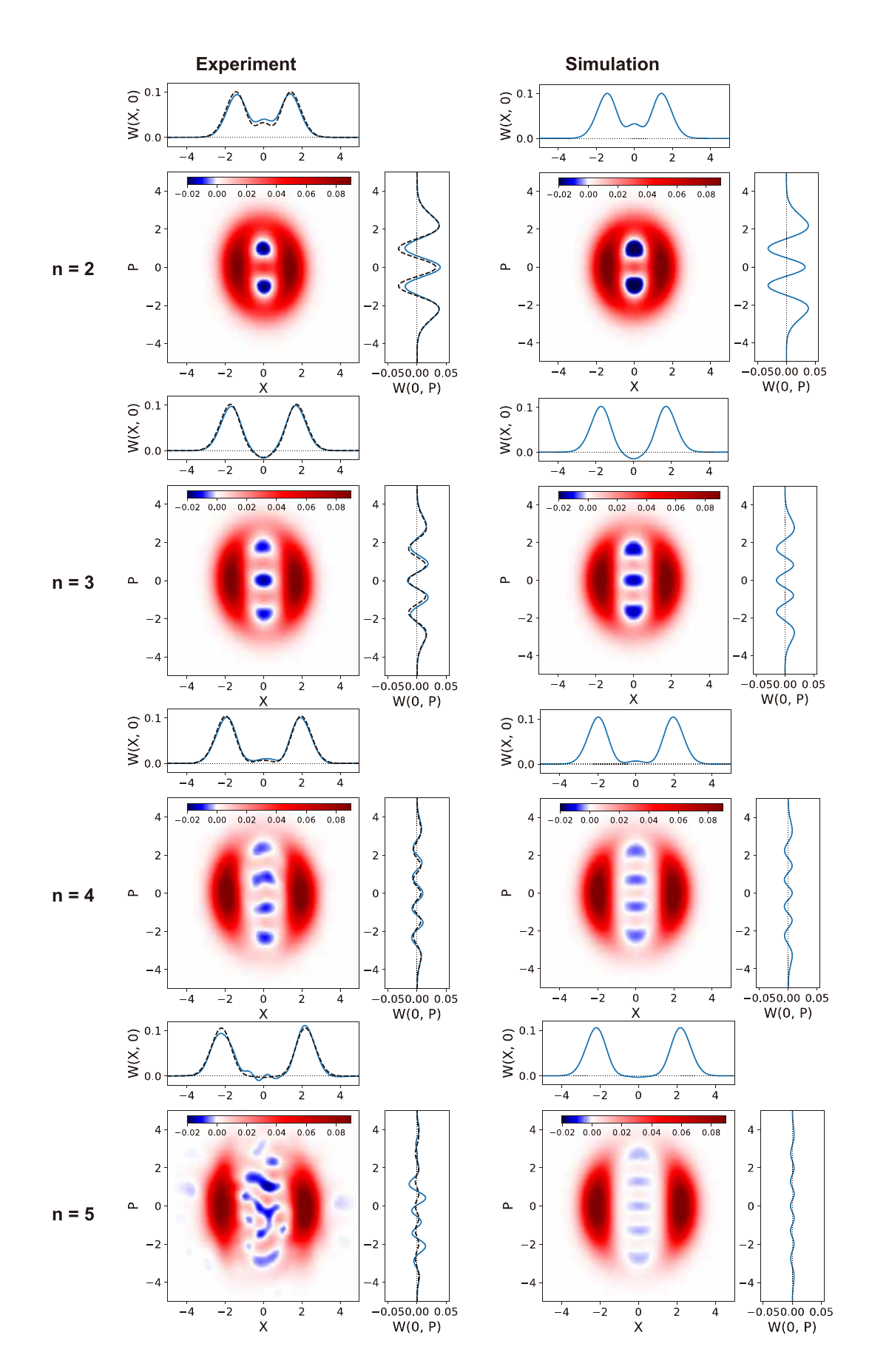}
\caption{Comparison between experimentally reconstructed and numerically simulated Wigner functions. For \(n\le4\), the experimental Wigner functions are identical to those shown in the main text. The \(n=5\) panel is included as preliminary qualitative data.}
\label{fig:SI_Wigner_compare}
\end{figure}

Figure~\ref{fig:SI_Wigner_compare} compares the experimentally reconstructed Wigner functions with the corresponding simulated Wigner functions. The overall structure, including the number and positions of the negative regions, is in close agreement.

\paragraph*{Preliminary data for \(n=5\).}
For the \(n=5\) conditioned events, the total number of acquired quadrature samples was insufficient to perform maximum-likelihood tomography with the same statistical confidence as for the \(n\le4\) data. The \(n=5\) state is intrinsically more complex and requires both a larger Hilbert-space truncation and substantially more samples for stable reconstruction.

Accordingly, we do not use the \(n=5\) data for quantitative claims in the main text. Instead, Fig.~\ref{fig:SI_Wigner_compare} includes the corresponding experimental and simulated Wigner functions for qualitative reference.

\FloatBarrier
\subsection*{S6. Estimation of the cat-state amplitude}

Here we describe in detail the procedure used to estimate the effective cat-state amplitude reported in the main text.

States generated by generalized photon subtraction (GPS) are well approximated as squeezed Schr\"odinger cat states. By numerically applying an appropriate inverse squeezing operation to the experimentally reconstructed state, one obtains a state close to an unsqueezed cat state of the form
\begin{align}
\ket{\mathrm{Cat}} \propto \ket{\alpha} \pm \ket{-\alpha}.
\end{align}

As discussed in the main text, the quadrature marginal distribution of a cat state along the \(x\) quadrature exhibits two distinct peaks corresponding to the coherent-state components. For an ideal unsqueezed cat state, each peak has the same width as the vacuum quadrature distribution, and the peak separation is
\begin{align}
\Delta x = 2\sqrt{2}\,\alpha
\end{align}
under the convention \(\hbar=1\).

\paragraph*{Operational definition of \(\alpha_{\mathrm{eff}}\).}
For each reconstructed state, we numerically applied an inverse squeezing operation until the widths of the two peaks in the \(\theta=0^\circ\) quadrature marginal matched that of the vacuum state. The effective cat-state amplitude was then extracted from the resulting peak separation,
\begin{align}
\alpha_{\mathrm{eff}}=\frac{\Delta x_{\mathrm{rescaled}}}{2\sqrt{2}}.
\end{align}

This procedure provides an operational estimate of the cat-state separation and does not rely on a global fidelity fit to an ideal cat-state model. Unless otherwise stated, no loss correction was applied in this procedure.

\paragraph*{Estimate before homodyne detection.}
For reference, we also report amplitude values corrected only for the loss in the homodyne-detection channel,
\begin{align}
\alpha_{\mathrm{in}}=\frac{\alpha_{\mathrm{eff}}}{\sqrt{\eta_{\mathrm{HD}}}},
\end{align}
where \(\eta_{\mathrm{HD}}\) is the independently measured homodyne efficiency. This quantity estimates the cat-state amplitude immediately before homodyne detection.

It is important to emphasize that this correction accounts only for losses associated with the homodyne detection system. No correction is applied for internal source loss, propagation loss, or inefficiency of the photon-number-resolving detector. Therefore, \(\alpha_{\mathrm{in}}\) should not be interpreted as a fully loss-compensated intrinsic amplitude.

\paragraph*{Relation to literature conventions.}
This definition differs from the amplitude estimate commonly used in previous optical cat-state experiments, in which the reconstructed state is first corrected for all known losses and the cat-state amplitude is then inferred from the squeezed cat state that maximizes the fidelity to the corrected state. Because that fully loss-corrected, model-dependent procedure deconvolves loss, it generally yields larger amplitudes. Direct numerical comparison between such literature values and the present \(\alpha_{\mathrm{eff}}\) is therefore not straightforward.

\subsection*{S7. Required cat-state amplitudes for fault-tolerant bosonic encodings}
\label{sec:supp_amplitudes}

This section summarizes the physical reasoning behind the cat-state amplitudes referenced in the main text, both for cat-code architectures and for adaptive breeding toward GKP-state generation.

\subsubsection*{S7.1 Cat-code architectures}

In two-component cat codes, the logical basis is spanned by the even and odd cat states $|\mathrm{Cat}_{\pm}\rangle \propto |\alpha\rangle \pm |-\alpha\rangle$. The bit-flip rate, corresponding to transitions between the two coherent-state components $|\alpha\rangle$ and $|-\alpha\rangle$, is exponentially suppressed in $\alpha^{2}$, while the phase-flip rate induced by photon loss grows approximately linearly in $\alpha^{2}$~\cite{Lund2008,Guillaud2019}. This asymmetry gives rise to the so-called biased-noise structure, which is a key resource for hardware-efficient fault-tolerant architectures. To reach a regime in which the bias is large enough to be useful for concatenated quantum error correction, sufficiently large amplitudes are required~\cite{Lund2008,Guillaud2019}. Higher-order rotationally symmetric cat codes (e.g., four-component cats) likewise demand sufficiently large $\alpha$ to render their logical basis states approximately orthogonal and to satisfy the approximate Knill--Laflamme conditions under photon loss~\cite{Cochrane1999, Mirrahimi2014, Leghtas2013, Hastrup2022, Bergmann2016, Li2017, Grimsmo2020}. In all of these settings, increasing $\alpha$ plays the role of increasing the effective code distance of the bosonic encoding.

\subsubsection*{S7.2 Adaptive breeding toward GKP states}

Adaptive breeding generates approximate GKP-like lattice states by interfering cat states (or cat-like intermediate states) on beam-splitter networks and applying Gaussian operations conditioned on homodyne-measurement outcomes~\cite{Weigand2018,Takase2024}. We consider a tree-type protocol with $m$ breeding steps, starting from squeezed cat states of (dimensionless) displacement amplitude $\alpha$, defined as the argument of the displacement operator $\hat{D}(\alpha)$.

Throughout this section we adopt the $\hbar = 1$ convention used in the main text, in which $\hat{x} = (\hat{a} + \hat{a}^{\dagger})/\sqrt{2}$ and $\mathrm{Var}(x_{\mathrm{vac}}) = 1/2$. In this convention, the coherent state $|\alpha\rangle$ has its Wigner-function peak at $x = \sqrt{2}\,\mathrm{Re}[\alpha]$, so the initial peak separation of an unsqueezed cat state along the relevant quadrature is
\begin{align}
\Delta x_{0} = 2\sqrt{2}\,\alpha,
\end{align}
consistent with the definition of $\alpha_{\mathrm{eff}}$ used in Sec.~S6.

Each breeding step introduces additional peaks between the existing ones, with the consequence that the spacing between neighboring peaks contracts by a factor of $\sqrt{2}$ per step. After $m$ steps the neighboring-peak spacing becomes
\begin{align}
\Delta x_{m} = \frac{2\sqrt{2}\,\alpha}{\sqrt{2^{\,m}}} = 2^{\,(3-m)/2}\,\alpha.
\label{eq:peak_spacing}
\end{align}
Concretely, the initial two-peak cat state with separation $2\sqrt{2}\,\alpha$ evolves into a three-peak lattice with neighboring spacing $2\alpha$ after $m = 1$, and into a five-peak lattice with neighboring spacing $\sqrt{2}\,\alpha$ after $m = 2$. For a two-step breeding the initial cat amplitude $\alpha$ therefore directly sets the final lattice spacing of the generated five-peak GKP-like state, up to the convention-fixed prefactor $\sqrt{2}$.

In the $\hbar = 1$ convention, the stabilizers of the symmetric qunaught GKP grid state \cite{Walshe2020}---the fully symmetric oscillator grid state that serves as an elementary resource for higher-level GKP codes and processing---are
\begin{align}
\hat{S}_{x} = \exp\left(i\sqrt{2\pi}\,\hat{p}\right),
\qquad
\hat{S}_{p} = \exp\left(-i\sqrt{2\pi}\,\hat{x}\right),
\end{align}
so that the logical state $|0\rangle_{L}$ has an $x$-quadrature comb of period $\sqrt{2\pi}$~\cite{Gottesman2001}. To match this lattice constant after $m = 2$ breeding steps, the initial cat amplitude must therefore satisfy
\begin{align}
\sqrt{2}\,\alpha = \sqrt{2\pi}
\quad \Longleftrightarrow \quad
\alpha \approx \sqrt{\pi} \approx 1.77.
\end{align}
The same dimensionless value $\alpha \approx 1.77$ corresponds to a peak separation $2\sqrt{2}\alpha \approx 5.0$ of the initial cat state in our $\hbar = 1$ convention, directly comparable to the experimentally extracted $\alpha_{\mathrm{eff}}$ defined in Sec.~S6.

Although a five-peak lattice does not by itself constitute a full GKP qubit, it already provides shift-error protection along one quadrature, which is the defining ingredient of GKP-type codes~\cite{Gottesman2001}. Larger lattices (and therefore stronger protection) require either more breeding steps or larger initial $\alpha$, both of which increase the experimental demands~\cite{Takase2024}.

\subsubsection*{S7.3 Relation to the present experiment}
The cat-state amplitude reported in the main text, $\alpha_{\mathrm{eff}} = 1.69$, obtained without any loss correction, lies in the regime of practical relevance for cat-code architectures and falls within ${\sim}5\%$ of the value $\alpha \approx 1.77$ required for a five-peak lattice with the GKP lattice constant. The platform therefore simultaneously enters the regime of practical relevance for cat-code bosonic encodings and approaches the amplitude required for adaptive breeding toward GKP-state generation, while preserving clear Wigner negativity in picosecond temporal modes.

\subsection*{S8. Breeding of the generated states}

To illustrate how the experimentally generated cat states behave under the breeding protocol introduced in Sec.~S7, we numerically applied tree-type breeding to two inputs: the experimentally reconstructed $n=4$ state (without loss correction) and its lossless counterpart obtained from the calibrated model of Sec.~S5, which reproduces the measured states with fidelity $F>0.99$. Each breeding step interferes two copies of the state on a balanced beam splitter and heralds on a homodyne measurement of the conjugate quadrature at outcome zero; this is the post-selected limit of the adaptive protocol, in which feedforward Gaussian operations relax the heralding condition and thereby raise the success probability~\cite{Takase2024, Weigand2018} without altering the peak-multiplication geometry.

The result is shown in Fig.~S9. In both cases the number of peaks in the position quadrature grows as $2\rightarrow3\rightarrow5$ over two breeding steps, and the neighbouring-peak spacing contracts by a factor of $\sqrt{2}$ per step, exactly as predicted by the tree protocol of Sec.~S7. For the experimental state the measured spacing evolves as $3.64\rightarrow2.70\rightarrow1.93$ (in $\hbar=1$ units), consistent with the $\sqrt{2}$ contraction.

The two cases differ in the fate of the Wigner negativity. In the lossless limit, breeding builds a two-dimensional GKP-like lattice whose negative regions are preserved ($W_{\min}=-0.25$, $-0.17$, $-0.17$ for the generated, one-step, and two-step states). For the experimentally reconstructed state, the same comb structure forms in the position quadrature, but because the state carries the full uncorrected loss the Wigner negativity decays as $W_{\min}=-0.0095(13)$, $-0.0043(10)$, $-0.0000(0)$, where the parenthetic uncertainties are bootstrap standard deviations; the two-step value is statistically indistinguishable from zero. The generated states therefore already supply the amplitude and lattice geometry required for breeding, while reaching a genuinely non-classical bred resource additionally requires the higher state purity that loss reduction would provide---consistent with the discussion in the main text.

\begin{figure}[t]
\centering
\includegraphics[width=\textwidth]{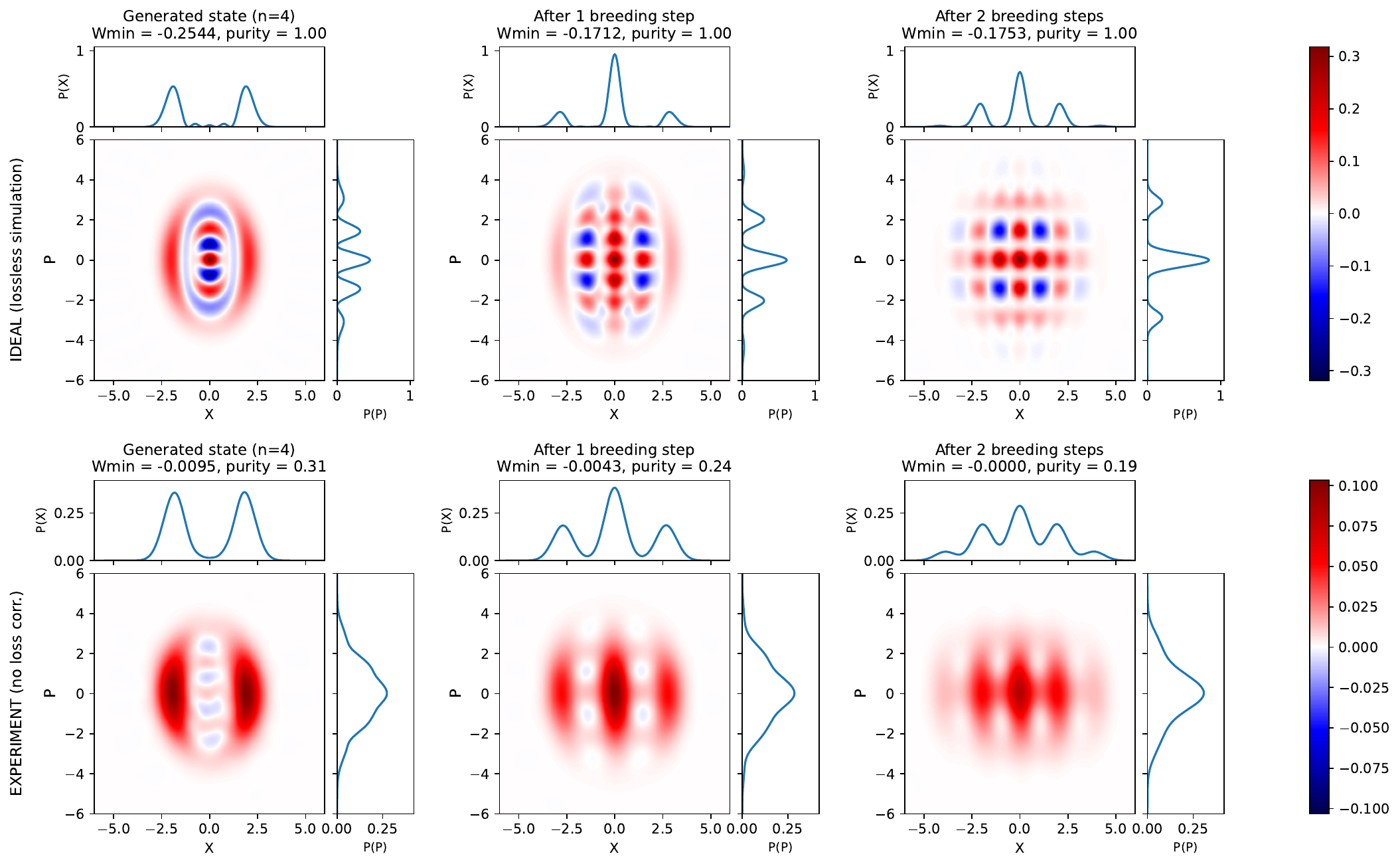}
\caption{Peak proliferation under tree-type breeding. Each panel shows the reconstructed Wigner function with its position-quadrature marginal $P(x)$ and $P(p)$. \textbf{Top:} lossless limit of the generated $n=4$ state (calibrated model of Sec.~S5). \textbf{Bottom:} experimentally reconstructed $n=4$ state without loss correction. \textbf{Columns:} generated state, after one breeding step, and after two breeding steps. The number of peaks increases $2\rightarrow3\rightarrow5$ in both cases. In the lossless limit the negative regions (blue) survive and form a two-dimensional GKP-like lattice; for the experimental state the comb forms in $x$ but the Wigner negativity is lost by the second step.}
\label{fig:S9}
\end{figure}

\clearpage
\bibliographystyle{Science}
\bibliography{scibib}

\end{document}